# Explainable Prediction of Economic Time Series Using IMFs and Neural Networks

**Pablo Hidalgo [1] , Julio E. Sandubete [1,\*], Agustín García-García[2]**

[1]   Faculty of Law, Business and Government, Universidad Francisco de Vitoria, 28223 Madrid, Spain.

[2]   Faculty of Economics and Business Studies, Universidad de Extremadura, 06006 Badajoz, Spain.

\*   Correspondence: je.sandubete@ufv.es

**Abstract:**

This study investigates the contribution of Intrinsic Mode Functions (IMFs) derived from economic time series to the predictive performance of neural network models, specifically Multilayer Perceptrons (MLP) and Long Short-Term Memory (LSTM) networks. To enhance interpretability, DeepSHAP is applied, which estimates the marginal contribution of each IMF while keeping the rest of the series intact. Results show that the last IMFs, representing long-term trends, are generally the most influential according to DeepSHAP, whereas high-frequency IMFs contribute less and may even introduce noise, as evidenced by improved metrics upon their removal. Differences between MLP and LSTM highlight the effect of model architecture on feature relevance distribution, with LSTM allocating importance more evenly across IMFs.

## 1. Introduction

Time series data offer a fundamental framework for modeling real-world phenomena as sequences of observations with temporal dependence. This temporal structure, together with the wide range of forecasting techniques developed to analyze it, constitutes a central pillar of the economic and financial literature, where generating accurate and well-justified forecasts is essential.

Until only a few decades ago, economic and financial time series were analyzed almost exclusively through econometric models. However, the rapid progress of digital technologies has encouraged the adoption of Artificial Intelligence (AI) methods, enabling high-frequency forecasting in which price-based variables are commonly used as inputs (Triebe et al., 2019).

Machine Learning (ML) models have achieved unprecedented predictive performance in recent years. Yet, their increasing complexity has introduced a major challenge: the difficulty of explaining and interpreting their results (Theissler et al., 2022). Interpretability is a foundational principle in economics, and the absence of transparency can hinder the extraction of meaningful insights, the detection of structural trends, and the formulation of robust policy or investment decisions.

A significant portion of existing AI-based decision-support systems has been developed as "black boxes," concealing the internal logic behind their predictions and raising practical as well as ethical concerns. This has driven a growing body of research aimed at clarifying how such models arrive at their outputs.

Early work in Explainable Artificial Intelligence (XAI) differentiated between interpretable and explainable models according to the principles they followed—such as feature importance, visual reasoning, or trend attribution—later grouping these approaches under broader methodological categories. Interpretable models include linear regression, decision trees, or attention-based architectures, whereas explainable models rely on perturbation-based, propagation-based, or visualization-based techniques.

XAI (Guidotti et al., 2019) therefore emerged as a research paradigm aimed at reducing opacity in AI systems by making their internal mechanisms understandable. Its goal is to map the abstract representations learned by a model into explanations that are meaningful to human users (Montavon et al., 2018). Although early developments in XAI were driven mainly by computer vision, recent years have witnessed growing interest in extending these methods to time series analysis (Benidis et al., 2023; Hewamalage et al., 2021), particularly in economics, where AI is increasingly central to forecasting and decision-making.

This study focuses specifically on explainability methods applied to neural networks—traditionally seen as "black box" models. Our objective is to extract interpretable insights from their predictive behavior in the context of univariate time series, an area where the literature remains limited. Whereas prior work on time-series explainability has primarily examined multivariate settings (Covert et al., 2020), emphasizing the identification of input features with the greatest predictive influence (Chandrashekar & Sahin, 2014), univariate data lack explicit features to analyze.

To address this limitation, we employ Empirical Mode Decomposition (EMD) to transform the original signal into a set of Intrinsic Mode Functions (IMFs), each representing a distinct oscillatory component. Treating these IMFs as standalone features enables us to determine which intrinsic elements of the series exert the greatest influence on model predictions. This approach enhances the interpretability of neural forecasting models while offering insights into the underlying economic dynamics encoded within the temporal structure of the data.

## 2. Related Work

### 2.1. XAI applied to Time Series Forecasting

In recent years, there has been a notable increase in techniques aimed at explaining machine learning models. Once the predictive performance of a model has been validated, incorporating interpretability mechanisms is often more efficient than redesigning the model entirely to make it inherently interpretable. Accordingly, several user-friendly libraries have been developed to support this goal.

One of the main challenges of this approach is that model-agnostic explanations are not intrinsically part of the underlying model, raising concerns about their fidelity. Therefore, a careful review of explainability techniques and their limitations is necessary — while the employed Artificial Neural Network (ANN) (Sundararajan et al., 2017) models themselves have already been widely validated, the focus here is on how explanations can shed light on important features in economic time series.

Much like interpretable modelling techniques, explainable models aim to assess the impact of each feature on the model's output. Two broad families dominate: **propagation-based** and **perturbation-based** methods. Propagation-based techniques track how each input feature's influence flows through the network layers to the output, while perturbation-based methods evaluate how modifications to input features affect the model's predictions.

### 2.2. Perturbation-Based Techniques

The perturbation approach estimates feature importance by modifying one feature at a time and comparing the altered prediction to the original (Breiman, 2001; Carta et al., 2022). Because this method does not depend on model internals, it is model-agnostic and applicable to any architecture. Classical examples include ablation, noise injection, or permutation importance. More advanced methods such as **SHAP** (Freeborough, 2022) and **LIME** (Ribeiro et al., 2016) have achieved widespread use. In financial time-series research, these techniques have been applied to recurrent neural networks (RNNs), demonstrating their complementary roles: ablation and integrated-gradient methods capture temporal and feature importance, while permutation importance assesses variable relevance independently of model internals (Cascarino et al., 2022). Findings from these studies suggest that price-related features are typically more influential than volume-related ones, and that recent observations carry higher predictive weight (Freeborough, 2022).

**SHAP** is an advanced, game-theory-inspired technique based on perturbations (Freeborough, 2022). It assigns each feature an importance value according to its marginal contribution to the model's output, by evaluating all possible combinations of presence or absence of features. SHAP ensures important properties such as local accuracy, consistency, and absence of bias (Lundberg et al., 2020). Depending on the type of model, different implementations may be used: **Kernel SHAP** for general models, or **TreeSHAP** optimized for tree-based algorithms like XGBoost. A key strength of SHAP is that it provides both **global and local** explanations, enabling a transparent interpretation of each variable's positive or negative contribution to individual predictions or overall model behavior (Weng et al., 2022).

**LIME** (Ribeiro et al., 2016), on the other hand, explains any black-box model by locally approximating it with a simpler, interpretable surrogate model — usually linear. It perturbs input data around a specific instance, generates synthetic samples, and fits a surrogate model whose coefficients indicate feature influence. Its main limitation is its inherently *local* scope: because it focuses on a small neighbourhood around one sample, its explanations do not generalize easily to global model behavior as SHAP does (Ghosh et al., 2023). This limitation is particularly acute in time series contexts, where temporal dependencies complicate the definition of a meaningful local perturbation neighborhood. For this reason, specialized adaptations like **TS-MULE** have been proposed to better handle temporal dependencies in time series settings (Schlegel et al., 2021).

## 3. Methodology

For this study, we employ a perturbation-based technique to enhance the explainability of deep learning models, using DeepSHAP to estimate the relevance of input features in the model's output. This approach enables us to assess how different components of the input contribute to the forecasting process, strengthening the interpretability of the results and offering a clearer understanding of the influence that each element exerts on the model's predictions.

Shapley values, rooted in cooperative game theory, offer a principled way to quantify the marginal contribution of each feature to a model's prediction, ensuring properties such as consistency and fairness in the attribution of importance (Lundberg & Lee, 2018). Formally, the Shapley value for a feature *i* is expressed as:

$$\phi_i = \sum_{S \subseteq N \setminus \{i\}} \frac{|S|! \, (|N| - |S| - 1)!}{|N|!} \, [f(S \cup \{i\}) - f(S)] \qquad (1)$$

where N denotes the full set of features and f(S) represents the model output when only the features in S are provided. These values satisfy local accuracy, missingness (features with no effect receive zero contribution), and consistency.

However, computing exact Shapley values becomes infeasible for large models, as the number of coalitions grows exponentially with |N|. To mitigate this, several approximation strategies have been proposed. One such approach is DeepLIFT (Shrikumar et al., 2017), which estimates the contribution of each input by comparing the network's activations to those produced under a reference or baseline input **x**′. DeepLIFT propagates relevance scores backward through the network using linear rules under the assumptions of independence among features and local linearity.

In general, DeepLIFT assigns a contribution $\mathbf{C}_{(\Delta \mathbf{x_j} \rightarrow \Delta \mathbf{t})}$ to each input $\mathbf{x_j}$ according to:

$$\mathbf{C}_{(\Delta \mathbf{x_j} \rightarrow \Delta \mathbf{t})} = \frac{\Delta \mathbf{t}}{\Delta \mathbf{x_j}} \Delta \mathbf{x_j} \qquad (2)$$

where $\Delta \mathbf{t} = \mathbf{f}(\mathbf{x}) - \mathbf{f}(\mathbf{x'})$ and $\Delta \mathbf{x_j} = \mathbf{x_j} - \mathbf{x'_j}$. If the baseline $\mathbf{x'}$ is chosen as the expected value of the inputs $\mathbf{E}[\mathbf{x}]$, DeepLIFT can be interpreted as an approximation to SHAP values under assumptions of independence and local linearity. Building on this connection, Lundberg & Lee (2017) introduced DeepSHAP, which combines the theoretical foundation of Shapley values with the efficiency of DeepLIFT. DeepSHAP treats DeepLIFT multipliers as compositional approximations of SHAP values and propagates them recursively to obtain an aggregate attribution across the full network.

The propagation rule for DeepSHAP can be written as:

$$\mathbf{m_i^{(l)}} = \sum_j \mathbf{m_j^{(l+1)}} \frac{\partial \mathbf{a_j^{(l+1)}}}{\partial \mathbf{a_i^{(l)}}} \qquad (3)$$

where $\mathbf{m_i^{(l)}}$ denotes the SHAP multiplier for node $\mathbf{i}$ at layer $\mathbf{l}$, and $\mathbf{a_j^{(l+1)}}$ is the activation in the subsequent layer.

In this way, DeepSHAP functions as a propagation-based technique that distributes relevance from the output layer back to the inputs, providing an interpretable representation of how each component influences the model's prediction.

### 3.1 Long Short-Term Memory (LSTM)

Long Short-Term Memory (LSTM) networks (Hochreiter & Schmidhuber, 1997) are a class of recurrent neural architectures specifically designed to capture temporal dependencies through memory cells that regulate the flow of information. Unlike feedforward architectures such as MLPs, LSTMs process sequences iteratively, preserving a hidden state that enables learning of both short- and long-term structure within time series.

In this work, the LSTM receives as input a sliding window of $\mathbf{N}$ past observations $(\mathbf{x_t}, \mathbf{x_{t-1}}, \dots, \mathbf{x_{t-N+1}})$ and generates a one-step-ahead prediction:

$$\hat{\mathbf{y}}_{t+1} = \text{LSTM}(\mathbf{x_t}, \mathbf{x_{t-1}}, \dots, \mathbf{x_{t-N+1}}) \qquad (4)$$

The model architecture consists of a single LSTM layer with ten memory units, followed by a dense output layer with one neuron corresponding to the next predicted value. The network is implemented in Python using Keras and TensorFlow, trained with the Adam optimizer (learning rate = 0.001) and a batch size of 64. Early stopping is employed to reduce overfitting. This configuration efficiently captures nonlinear and dynamic patterns characteristic of economic time series while maintaining training stability.

### 3.2 Multilayer Perceptron (MLP)

To model nonlinear relationships between past and future values, we also implement a Multilayer Perceptron (MLP), which takes as input a window of past observations $(\mathbf{x_t}, \mathbf{x_{t-1}}, \dots, \mathbf{x_{t-N+1}})$ and outputs a single step forecast:

$$\hat{\mathbf{y}}_{t+1} = \mathbf{f}(\mathbf{x_t}, \mathbf{x_{t-1}}, \dots, \mathbf{x_{t-N+1}}) \qquad (5)$$

where $\mathbf{f}(\cdot)$ represents the nonlinear function learned by the network. The proposed MLP includes one hidden layer with sixty-four neurons and an output layer with a single neuron. This configuration was selected after testing several architectures and obtaining consistently strong performance and stable training behavior. The model is built in TensorFlow/Keras and trained using the Adam optimizer for a maximum of 50 epochs, although convergence is typically reached earlier. The hidden layer uses a hyperbolic tangent activation to capture nonlinearities, while the output layer employs a linear activation suitable for continuous values. Classical references on the theoretical foundations of MLPs include Hornik et al. (1989) and Yi et al. (2023).

## 4. Experimentation

### 4.1 Dataset

This section outlines the dataset employed in the development and assessment of the proposed method. The analysis is based on economic time series, a type of data characterized by irregular dynamics, abrupt changes, and evolving patterns—features that make them an excellent benchmark for testing forecasting models. In addition, the economic domain places particular emphasis on interpretability, as transparent reasoning is essential for decisions derived from univariate predictive models.

For the experimental setup, the data must be split into two subsets: one used to fit the neural networks and another reserved for evaluating their performance. In this study, 75% of each series is allocated for training and the remaining 25% for validation. This partitioning is applied sequentially to the Intrinsic Mode Functions (IMFs) obtained from the EMD decomposition, rather than to the raw series.

An IMF is a nearly orthogonal oscillating component that must satisfy two main conditions: the number of extrema (maxima and minima) and zero-crossings must differ by a maximum of one, and the mean value of the envelope defined by the local maxima and local minima must be zero at any point. By decomposing our time series into several IMFs and a residual, the forecasting task can be simplified, as each IMF typically represents oscillations at a characteristic timescale, allowing for separate and often more accurate prediction of high-frequency (short-term) and low-frequency (long-term) patterns.

Regarding the primary dataset considered corresponds to NVIDIA's historical financial records. It contains 5,219 time-ordered observations spanning two decades, from May 2004 to May 2024, decomposed into 9 IMFs. The original series displays a clear long-term upward trajectory, with especially pronounced growth in the most recent period compared with the initial segment.

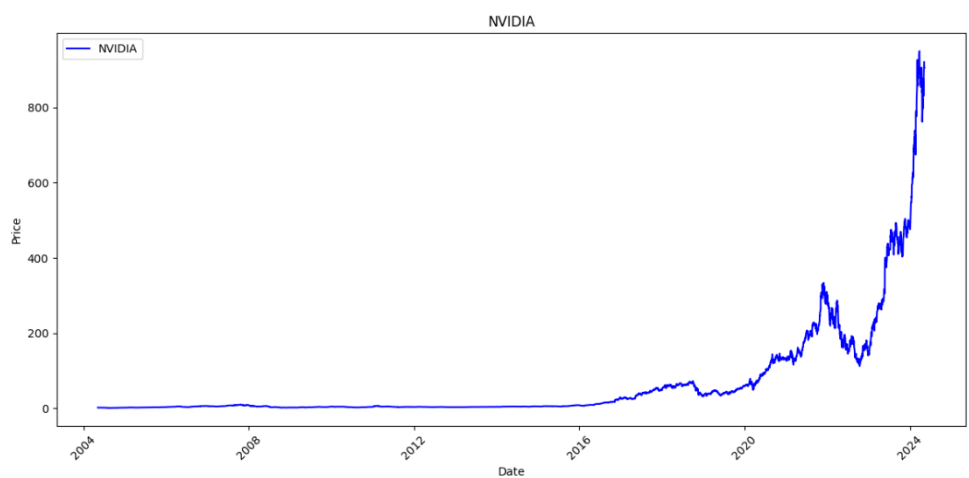

**Figure 1.** Graphical representation of the evolution of the NVIDIA time series from 2024 to 2024

A second dataset used in the analysis corresponds to Apple's stock price series, covering the same period—from May 2004 to 2024—with 5,218 data points and decomposed into 8 IMFs. Similar to the NVIDIA series, it presents a marked upward trajectory, especially in the later years. This behaviour introduces additional complexity for forecasting models: the distribution of values in the training segment differs markedly from that of the validation period. Consequently, the networks are required to extrapolate into a zone of accelerated growth, a scenario in which interpretability becomes particularly valuable for understanding how the model adjusts its forecasts under rapidly increasing trends.

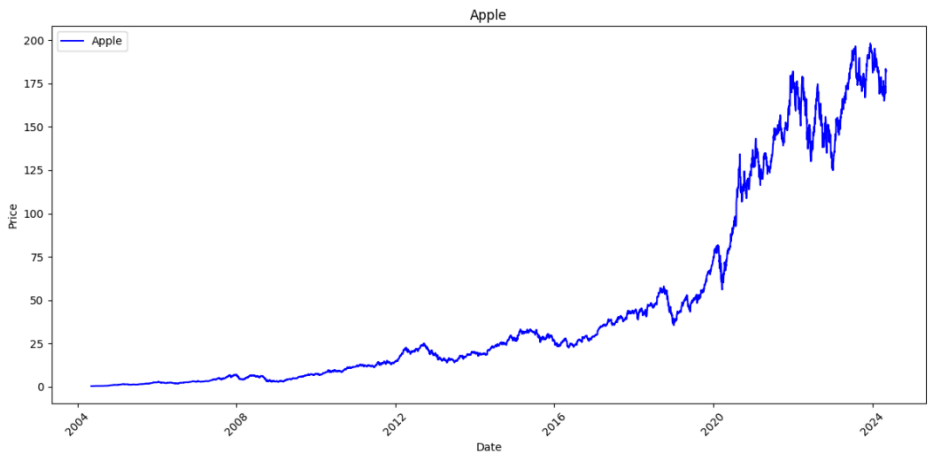

**Figure 2.** Graphical representation of the evolution of the APPLE time series from 2024 to 2024

Through error metrics, it is possible to quantitatively establish the accuracy with which the models make predictions, providing measurable methods to validate the fit. In this research, error metrics supported by the literature for time series prediction have been used (Botchkarev, 2019).

This error can be measured using the MSE and $R^2$ metrics, which evaluate the variation of the dependent variable based on the independent variable. The closer the value is to 1, the better the prediction as reflected by the $R^2$ metric. Meanwhile, MAPE and RMSE assess how accurate the model is in its predictions, so these values should be as low as possible. The use of $R^2$ has proven to be an effective technique for evaluating model performance, and when complemented with other metrics such as MAPE, MAE, and RMSE, it allows for a more complete analysis of the results and their comparison with previous studies, ensuring an objective view of them (Olawoyin & Chen, 2018).

## 5. Experimental Results and discussion

The time series were first decomposed into their corresponding IMFs, which served as inputs for the two neural network architectures considered in this study: an MLP and an LSTM network. To examine the interpretability of the forecasting process, we incorporated a perturbation-based explainability method, DeepSHAP, which provides an estimate of the average relevance of each IMF in the model's predictions.

The results obtained from this explainability analysis are presented for both model architectures, enabling a comparison of how each network processes the decomposed components of the series and whether similar interpretative patterns emerge across methods.

Figure 3 shows the MLP's reconstruction of the original NVIDIA time series, revealing a close alignment between predicted and actual values. The corresponding error metrics, summarized in Table 1, confirm the model's strong forecasting performance when all IMFs are included as inputs.

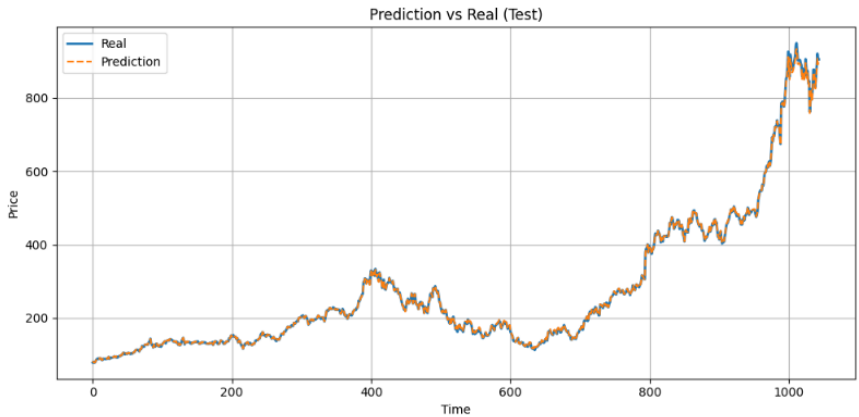

**Figure 3.** MLP results for Nvidia Time Series

**Table 1.** NVIDIA results.

| NVIDIA | |
| --- | --- |
| MSE | 13.605 |
| RMSE | 3.688 |
| MAE | 1.553 |
| R2 | 0.999 |

The next objective is to identify which of the IMFs processed by the model had the greatest impact. Figure 4 shows that IMF 9 has the highest contribution, with 70.4%. At a considerable distance, IMF 8 follows with 14.4% importance. This descending order continues, with IMF 4 showing a greater contribution than IMF 5, and IMF 1 exceeding IMF 2. Table 2 presents both the mean SHAP values and the corresponding importance percentage for each IMF in the prediction.

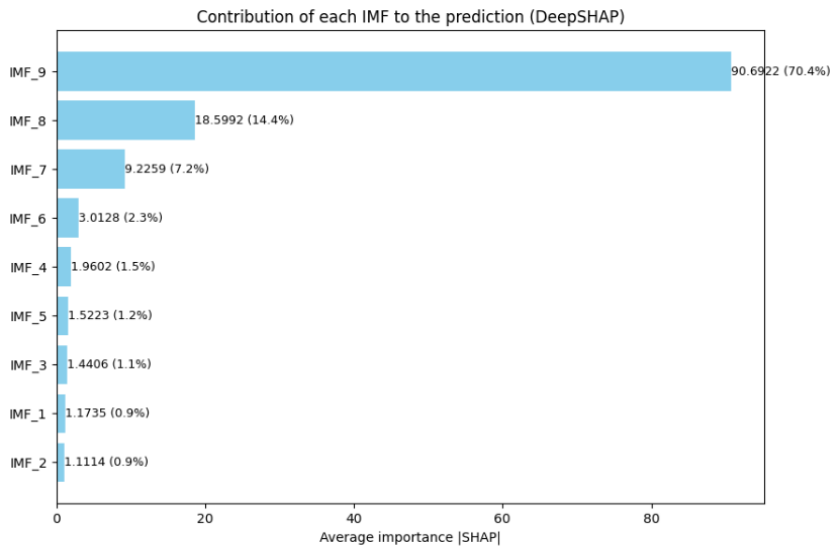

**Figure 4.** MLP DeepShap results for Nvidia Time Series

**Table 2.** NVIDIA DeepShap results

| IMF | Mean_Shap | Percent |
| --- | --- | --- |
| IMF_9 | 90.692 | 70.45% |
| IMF_8 | 18.599 | 14.45% |
| IMF_7 | 9.225 | 7.17% |
| IMF_6 | 3.012 | 2.34% |
| IMF_4 | 1.960 | 1.52% |
| IMF_5 | 1.522 | 1.18% |
| IMF_3 | 1.440 | 1.12% |
| IMF_1 | 1.173 | 0.91% |
| IMF_2 | 1.111 | 0.86% |

When conducting the same experiments using an LSTM network, the results remain consistent despite the architectural differences. Figure 5 illustrates the model's performance when all IMFs are provided as input to the LSTM, yielding error metrics that are satisfactory for this type of architecture, as reported in Table 3.

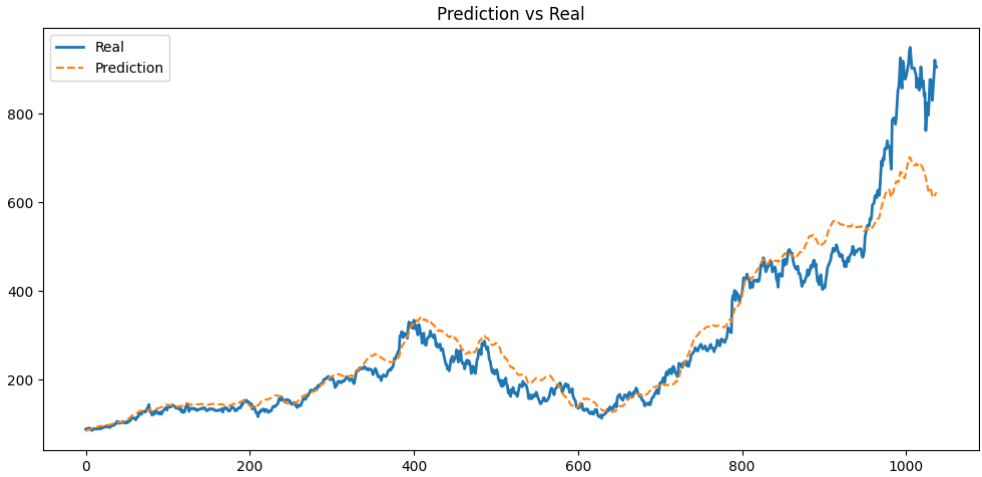

**Figure 5.** LSTM results for Nvidia Time Series

**Table 3.** NVIDIA results for LSTM

| | NVIDIA |
|---|---|
| MSE | 3481.235 |
| RMSE | 59.002 |
| MAE | 34.735 |
| R2 | 0.906 |

In this case, the IMF identified as most relevant by DeepSHAP continues to be the last one (Figure 6), although its percentage contribution is lower than that observed in the MLP model. This difference may stem from the distinct ways in which each architecture processes temporal dependencies: while the MLP captures relationships in a purely feed-forward manner, the LSTM incorporates sequential context, which can lead to a different distribution of importance across the IMFs. As shown in Table 4, the LSTM produces a more balanced importance profile than the MLP, suggesting that it distributes relevance more evenly among the different components of the decomposed series.

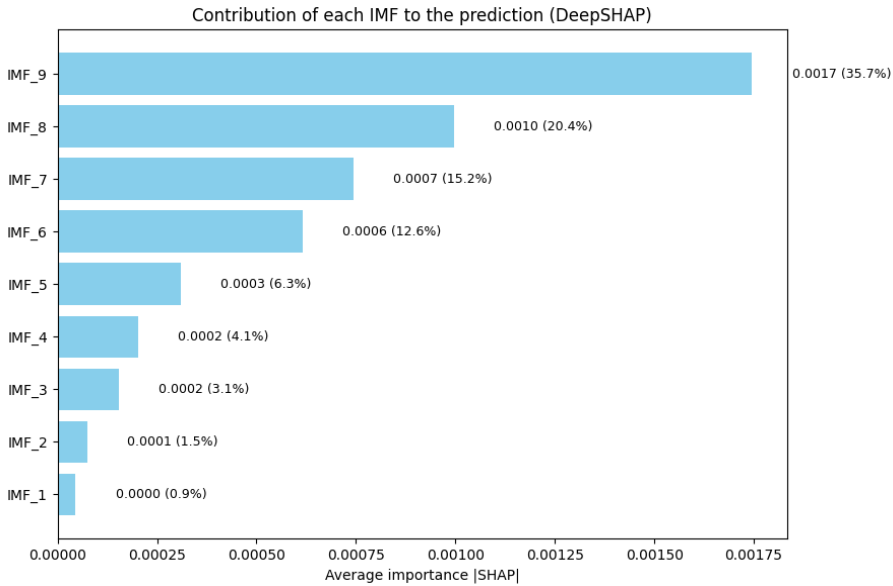

**Figure 6.** LSTM DeepShap results for Nvidia Time Series

**Table 4.** NVIDIA DeepShap LSTM results

| IMF | Mean_Shap | Percent |
|---|---|---|
| IMF_9 | 0.0017 | 35.73% |
| IMF_8 | 0.0010 | 20.38% |
| IMF_7 | 0.0007 | 15.21% |
| IMF_6 | 0.0006 | 12.61% |
| IMF_5 | 0.0003 | 6.32% |
| IMF_4 | 0.0002 | 4.14% |
| IMF_3 | 0.0001 | 3.15% |
| IMF_2 | 0.0000 | 1.53% |
| IMF_1 | 0.0000 | 0.93% |

When performing the same tests on the APPLE time series, similar results are observed. First, modeling the series with an MLP shows a strong fit when the resulting IMFs are used as inputs, as illustrated in Figure 7 and summarized in Table 5.

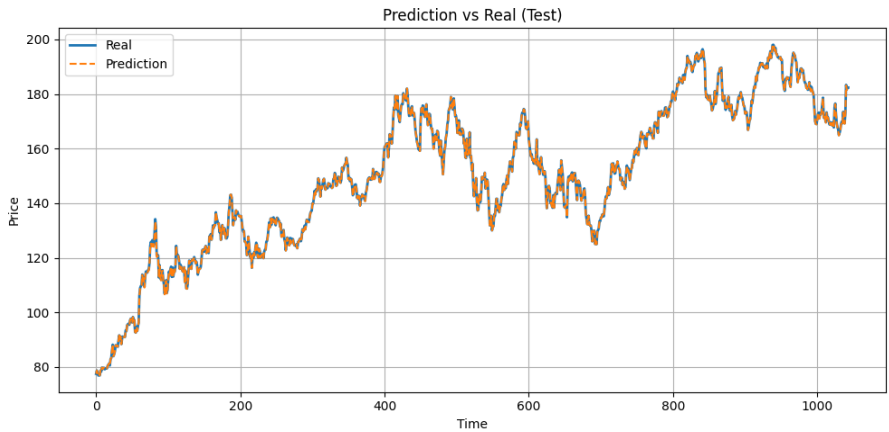

**Figure 7.** MLP results for APPLE Time Series

**Table 5.** APPLE results.

| | *NVIDIA* |
|---|---|
| MSE | 0.0149 |
| RMSE | 0.1221 |
| MAE | 0.0846 |
| MAPE | 0.06% |
| R2 | 0.999 |

Figure 8 illustrates that the greatest contribution to the model's predictions comes from the last IMF, number 8, which accounts for 78.8% of the overall importance, as shown in Table 6. The remaining IMFs follow a descending order of influence, with each successive component contributing progressively less to the model's output. This hierarchy highlights the dominant role of certain IMFs in capturing the main dynamics of the APPLE time series, while the lower-ranked IMFs provide complementary, fine-grained information that refines the predictions.

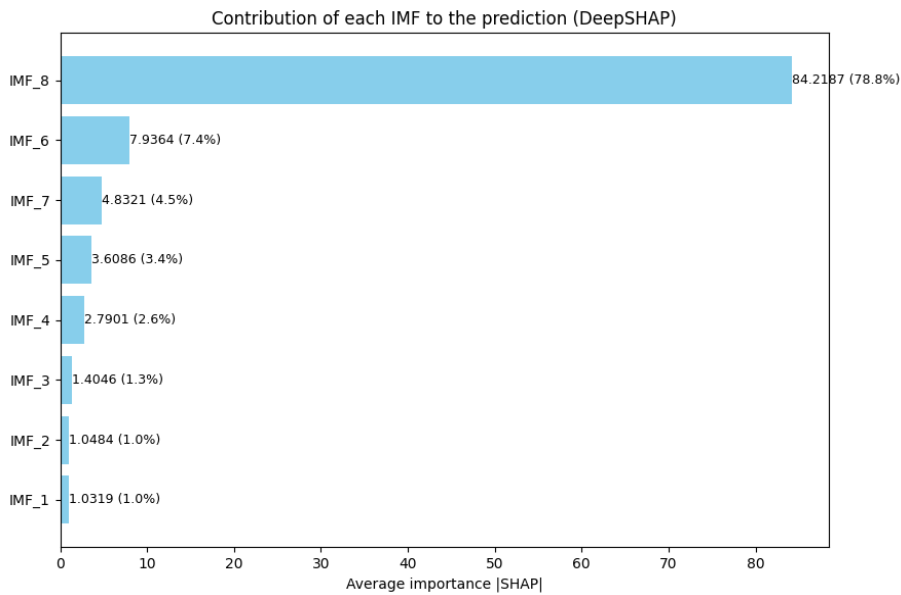

**Table 6.** APPLE DeepShap results

| *IMF* | *Mean_Shap* | *Percent* |
|---|---|---|
| IMF_8 | 84.218 | 78.80% |
| IMF_6 | 7.936 | 7.43% |
| IMF_7 | 4.832 | 4.52% |
| IMF_5 | 3.608 | 3.38% |
| IMF_4 | 2.790 | 2.61% |
| IMF_3 | 1.404 | 1.31% |
| IMF_2 | 1.048 | 0.98% |
| IMF_1 | 1.031 | 0.97% |

**Figure 8.** MLP DeepShap results for APPLE Time Series

Once again, the same analysis was conducted using an LSTM model, with the different IMFs provided as input for training. As illustrated in Figure 9, the LSTM demonstrates the ability to produce well-fitted predictions, achieving acceptable error metrics, as reported in Table 7.

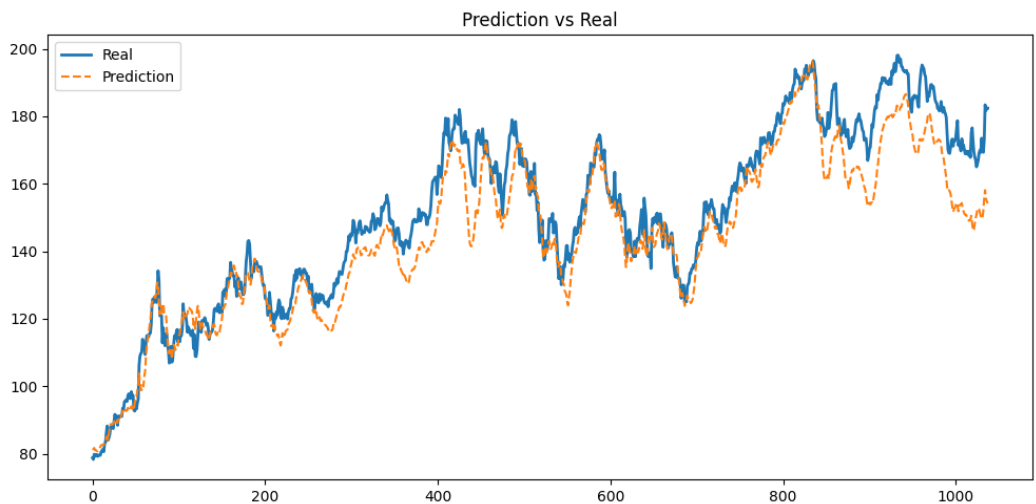

**Figure 9.** LSTM results for APPLE Time Series

**Table 7.** APPLE results for LSTM

|  | NVIDIA |
| --- | --- |
| MSE | 82.653 |
| RMSE | 9.091 |
| MAE | 7.042 |
| R2 | 0.884 |

For the APPLE time series, DeepSHAP highlights a pronounced contribution from the last IMF when using the LSTM model (Figure 10), similar to the pattern observed with the MLP. This IMF accounts for a significant 48.4% of the total contribution, as shown in Table 8. As with the other series, the LSTM model distributes importance more evenly across the different IMFs, providing a less extreme but still differentiated allocation compared to the MLP. This more balanced distribution can be attributed to the LSTM's ability to capture temporal dependencies, allowing it to integrate information from multiple IMFs over time rather than relying heavily on individual inputs.

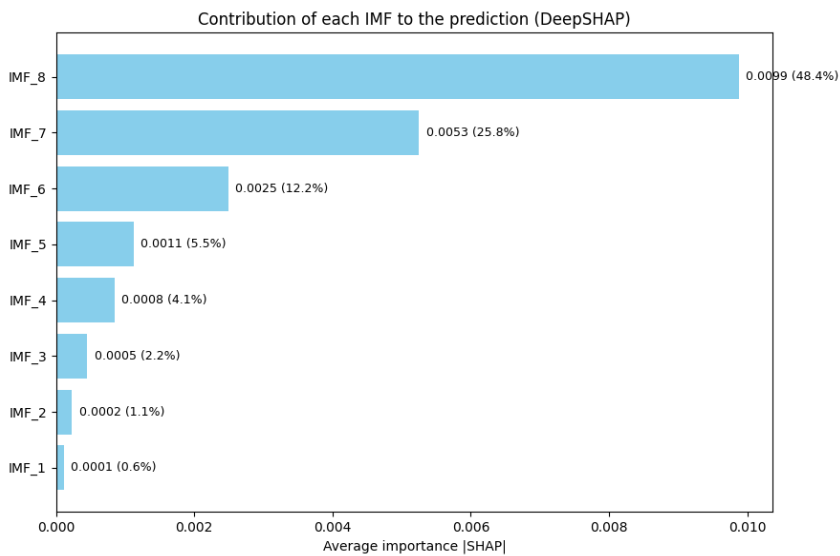

**Figure 10.** LSTM DeepShap results for APPLE Time Series

**Table 8.** APPLE DeepShap LSTM results

| IMF | Mean_Shap | Percent |
| --- | --- | --- |
| IMF_8 | 0.0098 | 48.40% |
| IMF_7 | 0.0052 | 25.75% |
| IMF_6 | 0.0025 | 12.25% |
| IMF_5 | 0.0011 | 5.53% |
| IMF_4 | 0.0008 | 4.15% |
| IMF_3 | 0.0004 | 2.23% |
| IMF_2 | 0.0002 | 1.14% |
| IMF_1 | 0.0001 | 0.56% |

## 6. Conclusions

In this study, the importance of the different IMFs obtained from economic time series was analyzed using two neural network architectures, MLP and LSTM, together with the interpretability technique DeepSHAP. The results show that this approach provides a comprehensive understanding of each IMF's relevance for prediction.

DeepSHAP estimates the marginal contribution of each IMF to the model's output, considering the influence of all other IMFs simultaneously. Examining individual IMFs reveals a consistent pattern: the last IMFs in the series, which represent the long-term trends of the signal, tend to be the most relevant according to DeepSHAP, as they capture the overall dynamics of the time series. Conversely, the initial IMFs reflect high-frequency components, oscillations, or noise, which may have a limited impact on prediction. This explains why the last IMFs are critical for modeling the global trend, while the first IMFs contribute less predictive information.

Comparing the architectures, both MLP and LSTM show generally consistent results, but with differences in how importance is distributed among the IMFs. The LSTM model, due to its memory capabilities and ability to capture temporal dependencies, distributes relevance more evenly across IMFs, showing less bias toward the high-trend components and assigning some importance to intermediate IMFs. In contrast, the MLP tends to concentrate importance on the last IMFs, reflecting its less adaptive nature with respect to the temporal structure of the series.

Overall, this study demonstrates that DeepSHAP is a robust and effective tool for interpreting the contribution of individual IMFs in neural network–based prediction models, highlighting how both model architecture and IMF characteristics—from noise to long-term trends—affect forecasting performance.